# A Self-assembled Fluoride-Water Cyclic Cluster of [F(H$_2$O)]$_4^{4-}$ in a Molecular Box

Md. Alamgir Hossain,*[1] Musabbir A. Saeed[1], Avijit Pramanik[1], Bryan M. Wong[2], Syed A. Haque[1], Douglas R. Powell[3]

[1]Department of Chemistry and Biochemistry, Jackson State University, Jackson, Mississippi 39217, United States

[2]Materials Chemistry Department, Sandia National Laboratories, Livermore, California 94551, United States

[3]Department of Chemistry and Biochemistry, University of Oklahoma, Norman, Oklahoma 73019, United States

Supporting Information

**ABSTRACT:** We present an unprecedented fluoride-water cyclic cluster of [F(H$_2$O)]$_4^{4-}$ assembled in a cuboid-shaped molecular box formed by two large macrocycles. Structural characterization reveals that the [F(H$_2$O)]$_4^{4-}$ is assembled by strong H-bonding interactions [OH···F = 2.684(3) to 2.724(3) Å], where a fluoride anion plays the topological role of a water molecule in the classical cyclic water octamer. The interaction of fluoride was further confirmed by $^{19}$F NMR and $^1$H NMR spectroscopies, indicating the encapsulation of the anionic species within the cavity in solution. High level DFT calculations and Bader topological analyses fully support the crystallographic results, demonstrating that the bonding arrangement in the fluoride-water cluster arises from the unique geometry of the host.

Ordered self-assembly in aqueous environments involves the spontaneous association of molecular species under certain conditions to generate highly structured aggregates stabilized by non-covalent interactions. In particular, the molecular interactions between fluoride anions and chemical receptors continue to be a topic of immense interest due to their significance in various environmental, biological, and health-related processes.[1-14] For example, the presence of fluoride anions in drinking water has been known to significantly impact human health, with direct effects on dental and skeletal fluorosis.[1-3] A fluoride-water adduct ([F(H$_2$O)]$^-$) also plays an important role in the stabilization of certain heme proteins such as *ferric sperm whale myoglobin*, where the fluoride anion is coordinated with one water (OH···F = 2.71 Å), a distal His64E7NE2 atom (NH···F = 2.74 Å) and the heme iron (Fe···F = 2.23 Å).[15] Herein we report a novel self-assembled [F(H$_2$O)]$_4^{4-}$ cluster formed within a cuboid-shaped molecular box provided by two parallel hosts, two water molecules, and two silicon hexafluorides. Within this self-contained molecular box, the unique geometry of the surrounding host provides an ideal microenvironment for assembling large hydrated guests.

Being the smallest member in the halide series, the fluoride anion is distinct from its congeners, displaying a high electronegativity and hydration energy.[16] This tiny anion has a high tendency to be hydrated instead of being isolated, making it even more challenging to bind with synthetic hosts in water. As recognized by Cametti and Rissanen, a "hydrated fluoride" instead of a "naked fluoride anion" has more relevance and importance within the anion recognition arena.[17] Although a hydrated fluoride has been the subject of extensive theoretical studies,[18-23] experimental evidence of fluoride-water cluster inside an enclosed cavity are remarkably lacking.[11,24-26] The recognition of hydrated fluoride was reported by Bowman-James and co-workers, describing the encapsulation of [F(H$_2$O)]$^-$ in an *m*-xylyl-based cryptand,[24] [F(H$_2$O)F]$^{2-}$ in a slightly expanded *p*-xylyl-based cryptand,[25] and [F(H$_2$O)$_4$]$^-$ in an amide-based tetrahedral host,[26] while the fluoride anion is tetrahedrally coordinated within a single molecule in each case. Recently, Ghosh and co-workers reported the formation of hydrated fluorides as [F$_4$(H$_2$O)$_{10}$]$^{4-}$ stabilized inside an amide based capsule.[11] However, such an assembled fluoride-water cyclic cluster within a closed nanocavity has not previously been observed.

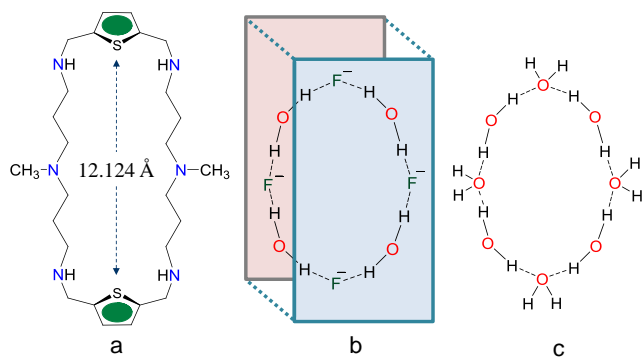

**Chart 1.** Schematics of (a) the receptor L, (b) the fluoride-water hybrid tetramer as [F(H$_2$O)]$_4^{4-}$ in a molecular box, and (c) the cyclic water octamer formed by four water molecules as H-bond donors and four water molecules as H-bond acceptors.

In 2001, Atwood and coworkers reported a cyclic (H$_2$O)$_8$ cluster, where four water molecules serve as two hydrogen bond donors and the remaining four as two acceptors.[27] We reported that NH$_2^+$ with H-bond donor groups and H$_2$PO$_4^-$ with both H-bond donor and acceptor groups can be successfully implemented in place of a water for the formation of an amine-water cyclic cluster [NH$_2^+$(H$_2$O)$_4$][28] and a dihydrogen phosphate octamer [(H$_2$PO$_4^-$)$_8$]$^{29}$, respectively. Here we report a self-assembled fluoride-water cyclic tetramer, [F(H$_2$O)]$_4^{4-}$, fully enclosed in a cuboid-type molecular box comprised by two large (length = 12.124 Å) parallel macrocycles. The fluoride anions and water molecules are hydrogen-bonded to each other in an alternating fashion within the fluoride-water hybrid cluster, where a fluoride anion plays the topological role of a water molecule in the classical cyclic water octamer (H$_2$O)$_8$ (chart 1).

The receptor L was synthesized from the reaction of an equimolar amount of *N*-methyl-3,3'-diaminodipropylamine and 2,5-thiophenedicarbaldehyde under high dilution conditions in CH$_3$OH followed by a reduction with NaBH$_4$.$^{30}$ The fluoride salt was obtained as a white powder after the addition of aqueous HF to L in CH$_3$OH in a Teflon vial. In order to avoid the possible contamination with SiF$_6^{2-}$ anions from a glass container,$^{24,25}$ attempts to grow crystals of the fluoride salt in a Teflon vial were unsuccessful. We therefore grew crystals in a glass vial from the slow evaporation of a CH$_3$OH/H$_2$O solution of the fluoride salt, providing colorless prism-shaped crystals as H$_6$[L]$^{6+}$·2SiF$_6^{2-}$·2F$^-$·10(H$_2$O).$^{31}$

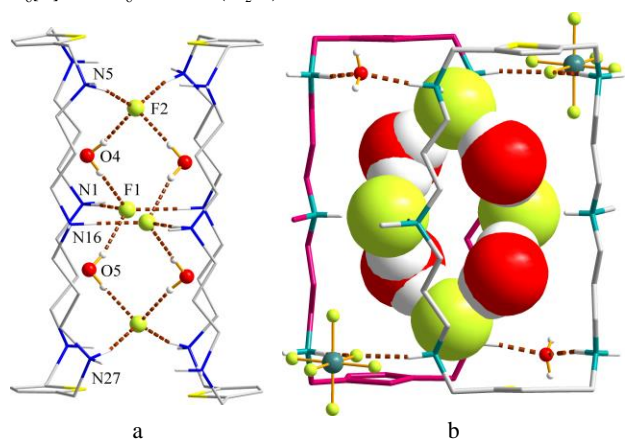

**Figure 1**. Crystal structure of the fluoride complex. (a) Side view showing the water-fluoride tetramer, [F(H$_2$O)]$_4^{4-}$, between the two macrocycles. (b) Space filling view of [F(H$_2$O)]$_4^{4-}$ in a molecular box formed by two parallel macrocycles, two water molecules, and two silicon hexafluorides.

**Table 1.** H-bonding parameters (Å, °) for the fluoride complex of L.

| D–H···F | H···F | D···F | ∠DHF |
|---|---|---|---|
| N1-H1···F1 | 1.71 | 2.628(3) | 166.7 |
| N5-H5A···F2 | 1.71 | 2.614(3) | 165.7 |
| N16-H16···F1$^i$ | 1.69 | 2.614(3) | 169.6 |
| N27-H27B···F2$^i$ | 1.68 | 2.587(3) | 167.0 |
| O4-H4OA···F1 | 1.94 | 2.724(3) | 166.7 |
| O4-H4OB···F2$^i$ | 1.89 | 2.684(3) | 172.5 |
| O4-H4OA···F1 | 1.94 | 2.724(3) | 166.7 |
| O5-H5OB···F2 | 1.94 | 2.724(3) | 167.1 |

Symmetry code: $^i$1-x+1, -y+1, -z+1

Structural analysis of the fluoride complex reveals that fluoride anions are assembled with water molecules in a highly ordered hydrogen bonding network to form a fluoride-water cyclic tetramer as [F(H$_2$O)]$_4^{4-}$ between the two hexaprotonated macrocycles. Each macrocycle adopts a rectangular shape, and the two thiophene rings in a macrocyclic unit are oriented parallel to each other with the distance of Ar$_{centroid}$···Ar$_{centroid}$ of 1.2124 Å. Figure 1a shows that each fluoride is coordinated with two NH groups from two macrocycles and with two water molecules, completing a fluoride-water cyclic tetramer ([F(H$_2$O)]$_4^{4-}$) between the two parallel macrocycles. Each fluoride is strongly bonded with four H-bond-donors in a tetrahedral coordination geometry, as previously observed in Lehn's BISTREN for naked fluoride.$^{32}$ The hydrogen bonding interactions of (NH···F) and (OH···F) range from 2.587(3) to 2.628(3) Å and 2.684(3) to 2.724(3) Å, respectively (Table 1). The corresponding distances are comparable to that reported for the fluoride complex of *p*-xylyl cryptand (NH···F = 2.65 Å),$^{25}$ and for the fluoride-water cluster ([F(H$_2$O)]$^-$) inside the *ferric sperm whale myoglobin* (NH···F = 2.74 Å and OH···F = 2.71 Å).$^{15}$ In the cyclic cluster, four fluorides act as double H-bond acceptors and four water molecules as double H-bond donors. The bonding patterns are surprisingly similar to that for the cyclic (H$_2$O)$_8$ cluster reported by Atwood and co-workers,$^{23}$ with four water molecules as donors and four water molecules as acceptors (see, Scheme 1c).

In addition to the formation of a fluoride-water cyclic cluster, two macrocycles are further connected with two water molecules and two silicon hexafluorides via NH protons (N···O = 2.698(4) and 2.685(5) Å; NH···F = 2.708(4) to 3.004(4) Å), respectively, resulting in a cuboid type molecular box. Clearly, both the silicon hexafluorides and water molecules play an important structural role in holding the two macrocycles together through hydrogen bonding interactions. The space-filling view of this tetramer (Figure 1b), illustrates the compact arrangement of water-fluoride cyclic tetramer stabilized in the molecular box. It is worth mentioning that all fluoride anions in the complex are fully utilized for the cluster formation in the solid state.

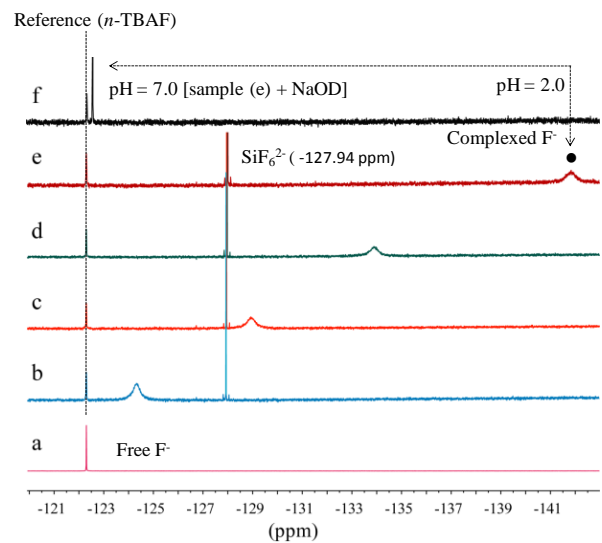

**Figure 2**. $^{19}$F NMR spectra of *n*-Bu$_4$N$^+$F$^-$ in D$_2$O recorded at room temperature: (a) Free *n*-Bu$_4$N$^+$F$^-$ (δ$_F$ = -122.3 ppm), (b) *n*-Bu$_4$N$^+$F$^-$ + 1.5 equiv. of H$_6$L(Ts)$_6$ (δ$_F$ = -124.3 ppm), (c) *n*-Bu$_4$N$^+$F$^-$ + 2.5 equiv. of H$_6$L(Ts)$_6$ (δ$_F$ = -128.9 ppm), (d) *n*-Bu$_4$N$^+$F$^-$ + 3.5 equiv. of H$_6$L(Ts)$_6$ (δ$_F$ = -133.8 ppm) and (e) *n*-Bu$_4$N$^+$F$^-$ + 5.0 equiv. of H$_6$L(Ts)$_6$ (δ$_F$ = -141.8 ppm), and (f) after adding NaOD to the sample (e) [pH = 7.0].

$^{19}$F NMR spectroscopy was used to characterize the chemical environment of fluoride anions in the presence of H$_6$L(Ts)$_6$ in water. For this purpose, the $^{19}$F NMR spectra were recorded for the fluoride solution (5 mM *n*-tetrabutylammonium fluoride in D$_2$O) before and after the addition of the host (50 mM in D$_2$O)

at pH = 2.0. In order to make a direct comparison, the same solution of n-tetrabutylammonium fluoride (5 mM in $D_2O$) was used as an external reference in a sealed capillary tube, and placed in the NMR tube. As clearly shown in Figure 3, the signal at $\delta_F$ = -122.3 ppm assigned to the unbound fluoride significantly shifts upfield to $\delta_F$ = -141.8 ppm ($\Delta\delta$ = 29.5 ppm) due to the addition of five equivalents of $H_6L(Ts)_6$, indicating the encapsulation of fluoride in the cavity. Furthermore, a new peak emerges at $\delta_F$ = -127.9 ppm, which remains unchanged during the titration process (Figure 2b-e). This peak could be assigned to a silicon hexafluoride ($SiF_6^{2-}$)[33,34] and is formed due to the addition of TBAF at low pH. The formation of $SiF_6^{2-}$ is quite common, particularly in the presence of HF or fluoride salts,[35] and is also consistent with the results of our crystallographic data. However, upon the addition of NaOD to the solution of TBAF containing five equivalents of the host (Figure 2e), two significant changes were observed: the peak at $\delta_F$ = -127.9 ppm disappeared, and the peak at $\delta_F$ = 133.8 ppm shifted downfield to $\delta_F$ = -122.0 ppm. The disappearance of the peak at $\delta_F$ = -127.9 ppm for $SiF_6^{2-}$ could be due to the reaction of $SiF_6^{2-}$ + NaOH →NaF + $SiO_2$ + $H_2O$ (used in water fluoridation).[36] However, the huge upfield shift close to the unbound fluoride ($\delta_F$ = -122.3 ppm) could be the result of the decomplexation of fluoride at higher pH (~ 7.0), as expected.

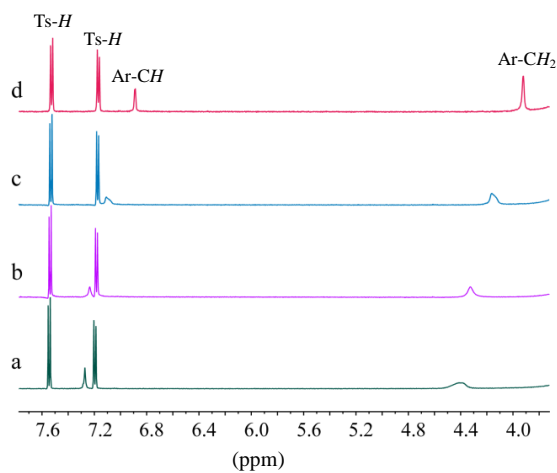

**Figure 3**. Partial $^1H$ NMR spectra of $H_6[L](Ts)_6$ (2 mM) with an increasing amount TBAF (20 mM) in DMSO-$d_6$: a = $H_6[L](Ts)_6$; b = $H_6[L](Ts)_6$ + 1.0 equiv. of TBAF; c = $H_6[L](Ts)_6$ +2.0 equiv. of TBAF; d = $H_6[L](Ts)_6$ +4.0 equiv. of TBAF.

**Table 2.** Binding constants (in log K) for the receptor $H_6L(Ts)_6$ with anions in DMSO-$d_6$ at 25 °C.

| $F^-$ | $Cl^-$ | $Br^-$ | $I^-$ | $NO_3^-$ | $ClO_4^-$ | $HSO_4^-$ | $H_2PO_4^-$ |
|---|---|---|---|---|---|---|---|
| 1.95 3.47[a] | 2.13 | 1.94 | < 1 | 1.72 | < 1 | 2.21 | [b] |

[a]The stepwise binding constant for $K_{11}$ and $K_{12}$ for the 1:1 and 1:2 (L:A) binding, respectively. [b]NMR titration was hampered due to the precipitation upon the addition of the anion.

$^1H$ NMR titration studies were performed to evaluate the binding affinity of $H_6L(Ts)_6$ for various anions of n-$Bu_4N^+$ salts in DMSO-$d_6$. As shown in Figure 3, the addition of n-$Bu_4NF$ to the host resulted in an upfield shift of both ArH and $CH_2$ resonances of the macrocycle (Figure 3). Our results show that the receptor exhibits the strongest interaction for $F^-$ over other anions (Table 2), providing the best fit for 1:2 (L:A) binding mode.[37] The 1:2 stoichiometry, as confirmed by Job plot analysis, is in agreement with the crystallographic results.

In order to quantitatively understand the unique assembly of the fluoride-water cluster within the molecular box, density functional theory (DFT) calculations were performed on the enclosed $[F(H_2O)]_4^{4-}$ species between the two macrocycles in (a) the absence and (b) the presence of linking groups ($H_2O$ and $SiF_6^{2-}$). All quantum chemical calculations were carried out with the M06-2X hybrid functional, which we have previously shown to accurately predict the binding energies of water and ions within large molecular systems.[28,38] All molecular geometries were completely optimized without constraints at the M06-2X/6-31G(d,p) level of theory in the presence of a polarizable continuum model (PCM) solvent model to approximate an aqueous environment (dielectric constant = 78.4). From the DFT optimized geometries, we calculated the cohesive energies (=$E_{anions}$+2×$E_{ligand}$−$E_{total\_complex}$) of both complexes. While both systems provided stable fluoride-water clusters (see, supporting information for optimized geometries and total energies), it is interesting to note that the complex with the $[F(H_2O)]_4^{4-}$ species in the molecular box formed by two macrocycles, two linking $H_2O$ groups, and two linking $SiF_6^{2-}$ is energetically more favorable (25.4 kcal/mol) than the complex without the linking groups (18.6 kcal/mol). The bonding patterns in the complex is surprisingly similar (Figure 4) to that observed in the X-ray data (*vide supra*).

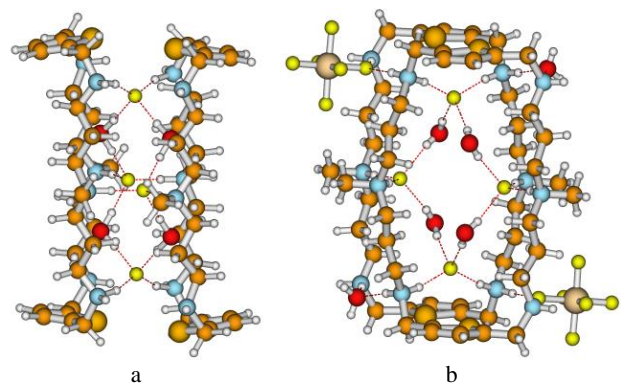

**Figure 4.** Optimized DFT geometry of $[F(H_2O)]_4^{4-}$ (a) between the two hosts, and (b) in the molecular box formed by two hosts, two water molecules and two silicon hexafluorides at the M06-2X/6-31G(d,p) level of theory.

To characterize the unique topological properties of the electron density distribution even further, we also used Bader's Quantum Theory of Atoms-In-Molecules (QTAIM) approach to calculate bond critical points (BCPs) throughout the complex. Within this approach, the presence of a BCP as well as a positive value of the Laplacian of the density ($\nabla^2\rho$) at the BCP gives a quantitative measure of the strength of a particular bond. Within our enclosed water cluster, we found that all of the hydrogen bonds in both the N-H···F and O-H···F groups are quite strong with large positive values. However, our QTAIM analysis indicates that the N-H···F bonds are considerably stronger with $\nabla^2\rho$= 0.20, while all the other O-H···F bonds are weaker with $\nabla^2\rho$

< 0.13. The relative strengths of these hydrogen bonds reflect the different electronegativities of the heteroatom (which are also in agreement with our QTAIM charge population analysis) and highlight the important role of the relative orientation of the N-H functional groups in the box.

In conclusion, we have presented the assembly of a fluoride-water cyclic tetramer of $[F(H_2O)]_4^{4-}$ within a molecular box comprised of a pair of large macrocyclic frameworks. Each macrocycle is preorganized in such a way to donate directional H-bonds for fluoride anions and to provide a precise space for water molecules, resulting in the formation of a fluoride-water hybrid cluster. Four fluoride anions and four water molecules are alternatively assembled to form an octameric cycle, where fluorides play the topological role of water molecules in the classical octameric water cluster $(H_2O)_8$.[27] The assembly of the stable fluoride-water cluster is fully supported by high level DFT calculations and Bader topological analyses, demonstrating that the unique assembly of the fluoride-water cluster results from the precisely positioned binding sites in the macrocycles as well as the linking water and hexafluoride species. In addition to presenting an exceptional example of a highly organized anion-water cyclic tetramer of $[F(H_2O)]_4^{4-}$ within the large cavity, this modular approach based on a large host leads to promising new types of self-assembled structures and a step towards the understanding of complex aqueous phase environments of an anion. Such self-assembled clusters inside a closed cavity may be useful for the generation of novel functional materials. Further studies on the assembly of other anion clusters with this and its analogues are currently in progress.

## ASSOCIATED CONTENT

Crystallographic data in CIF format. $^1H$, 2D, and $^{13}C$ NMR, and mass spectra, NMR titration spectra, additional structures of the fluoride complex, optimized geometries, and electronic energies. This material is available free of charge via the Internet at http://pubs.acs.org.

## AUTHOR INFORMATION

**Corresponding Author**

alamgir.hossain@jsums.edu.

## ACKNOWLEDGMENT

The National Science Foundation is acknowledged for a CAREER award (CHE-1056927) to MAH. This work was supported by the National Institutes of Health (G12RR013459). The NMR instrument used for this work was funded by the National Science Foundation (CHE-0821357). The authors thank the National Science Foundation (CHE-0130835) and the University of Oklahoma for funds to acquire the diffractometer used in this work.

(36) Aigueperse, J.; Mollard, P.; Devilliers, D.; Chemla, M.; Faron, R.; Romano, R.; Cuer, J. P. *"Fluorine Compounds, Inorganic"* in *Ullmann's Encyclopedia of Industrial Chemistry*, Wiley-VCH, Weinheim, 2005.

(37) Hynes, M. J. *J. Chem. Soc. Dalton Trans.* **1993**, 311–312.

(38) Saeed, M. A.; Pramanik, A.; Wong, B. M.; Haque, S. A.; Powell, D. R.; Chand, D. K.; Hossain, M. A. *Chem. Commun.* **2012**, DOI: 10.1039/C2CC30767G.

A self-assembled fluoride-water cyclic tetramer $(F(H_2O))_4^{4-}$ is formed in a cuboid shaped molecular box comprised of two parallel rectangular-shaped hosts.

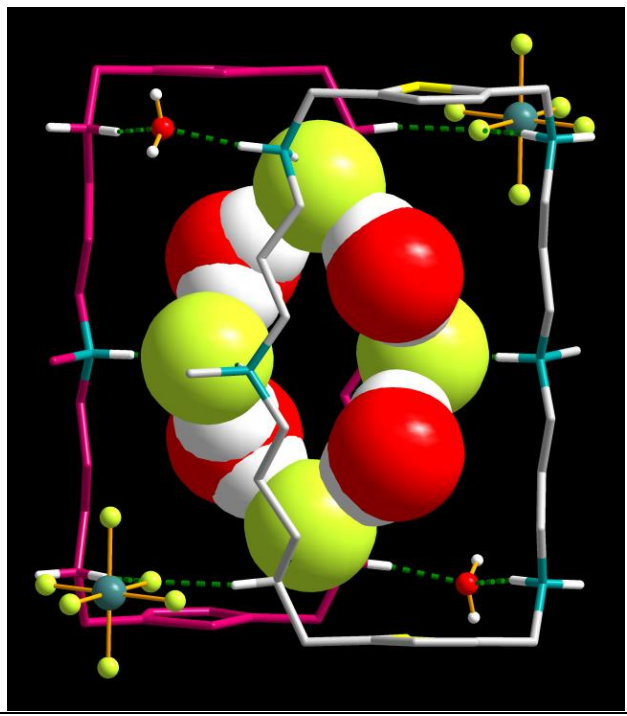